\documentclass[twocolumn]{aastex631}
\usepackage{array}
\usepackage{amsmath}
\usepackage{xspace}

\usepackage{graphicx,subfigure,comment}
\newcommand{\lwp}{{\sc LWP}\xspace}
\newcommand{\bilby}{{\sc Bilby}\xspace}

\begin{document}

\title{A Realistic Projection for Constraining Neutron Star Equation of State with the LIGO-Virgo-KAGRA Detector Network in the A+ Era}


\author[0009-0001-9946-9700]{Alexis Boudon}
\affiliation{Univ Lyon, Univ Claude Bernard Lyon 1, CNRS/IN2P3, IP2I Lyon, UMR 5822, F-69622, Villeurbanne, France}
\email{a.boudon@ip2i.in2p3.fr}

\author[0000-0001-6339-1537]{Hong Qi}
\affiliation{School of Mathematical Sciences, Queen Mary University of London, 327 Mile End Road, London, E14NS, United Kingdom} 
\affiliation{Department of Physics and Astronomy,   Louisiana State University, Baton Rouge, LA 70803, USA}
\email{hong.qi@ligo.org}

\author[0000-0002-5019-2720]{Jean-Francois Coupechoux}
\affiliation{Univ Lyon, Univ Claude Bernard Lyon 1, CNRS/IN2P3, IP2I Lyon, UMR 5822, F-69622, Villeurbanne, France}

\author[0000-0002-8457-1964]{Philippe Landry}
\affiliation{Canadian Institute for Theoretical Astrophysics, University of Toronto, Toronto, ON M5S 3H8, Canada}
\affiliation{Perimeter Institute for Theoretical Physics, Waterloo, Ontario N2L 2Y5, Canada}

\author[0000-0003-0885-824X]{Viola Sordini}
\affiliation{Univ Lyon, Univ Claude Bernard Lyon 1, CNRS/IN2P3, IP2I Lyon, UMR 5822, F-69622, Villeurbanne, France}

\begin{abstract}
The LIGO-Virgo-KAGRA network in the upcoming A+ era with upgrades of both Advanced LIGO (aLIGO+) and Advanced Virgo (AdV+) will enable more frequent and precise observations of binary neutron star (BNS) mergers, improving the constraints on neutron star equation of state (EOS). In this study, we applied reduced order quadrature techniques for full parameter estimation of 3,000 simulated gravitational wave signals from BNS mergers at A+ sensitivity following three EOS models—\textsc{hqc18}, \textsc{sly230a}, and \textsc{mpa1}. We found that tidal deformability tend to be overestimated at higher mass and underestimated at lower mass. We postprocessed the parameter estimation results to present our EOS recovery accuracies, identify biases within EOS constraints and their causes, and quantify the needed corrections.

\end{abstract}

\section{Introduction} \label{sec:intro}

Seven years on from the first observation of a binary neutron star (BNS) merger with gravitational waves (GWs), GW170817~\citep{GW170817,GW170817_mma,GW170817_source} remains the only GW transient that is appreciably informative about the uncertain structure and makeup of neutron star (NS) matter. The measurement of the BNS tidal deformability parameter $\tilde{\Lambda}$~\citep{FlanaganHinderer2008,WadeCreighton2014}from the tidal phasing of the inspiral gravitational waveform has been translated into constraints on the NS radius and equation of state (EOS)~\citep{DeFinstad2018,GW170817_eos,GW170817_models}. These constraints have since been refined~\citep{CapanoTews2020,LegredChatziioannou2021,HuthPang2022,KoehnRose2024} by combining information from radio~\citep{AntoniadisFreire2013,FonsecaCromartie2021} and X-ray~\citep{MillerLamb2019,MillerLamb2021,RaaijmakersRiley2019,SalmiChoudhury2024} observations of pulsars, data from nuclear experiments~\citep{LeFevreLeifels2016,RussottoGannon2016,AdhikariAlbataineh2021}, and predictions from nuclear theory~\citep{LynnTews2016,DrischlerHebeler2017}, while recent works have explored the impact of dynamical tidal effects and advanced Bayesian methods for EOS constraints in future GW observations \citep{Pradhan:2023xtq, Ghosh:2024cwc}.

Nonetheless, significant uncertainties exist in our understanding of NS matter and its equation of state.

The current dearth of revelatory GW observations of NSs is attributable to two facts. First, the astrophysical rate of BNS mergers is much lower~\citep{O3bPop} than estimated in the immediate aftermath of GW170817~\citep{GW170817}: despite 11 months of subsequent observing time by the LIGO-Virgo-KAGRA (LVK) network~\citep{Aasi2015,AcerneseAgathos2015,AkutsuAndo2021} with improved detector sensitivity during the O3 observing run, only one additional BNS merger, GW190425~\citep{GW190425}, has been observed to date. Second, GW190425's source was so distant, and so unusually massive, that no useful tidal information was extractable from the signal~\citep{GW190425}. Similarly, the LVK's existing NS--black-hole discoveries~\citep{NSBHs,GW230529}---while more numerous than the BNS detections---have very little to say about NS matter, as they are unlikely to have undergone tidal disruption prior to merger owing to their asymmetric mass ratios and small primary spins~\citep{BiscoveanuLandry2023,Sarin:2023tgv,GW230529}. 

After its ongoing O4 observing run concludes, the LVK network will undergo further upgrades to bring its LIGO detectors to A+ sensitivity~\citep{ObservingScenarios} and its overall BNS range~\citep{ChenHolz2021} to 240-345 Mpc~\citep{LVKobsplan}. Constraints on the neutron star EOS at Advanced LIGO/Virgo design sensitivity have been studied, such as \cite{Agathos:2015uaa} with 200 BNSs and two parameter power series for the tidal deformability versus mass relation, \cite{HernandezVivanco:2019vvk} with 40 BNSs and piecewise polytrope and spectral EOSs, and \cite{Landry:2020vaw} with 5 BNSs and nonparametric EOSs; however, these studies used outdated estimation of A+ sensitivity, simulations at smaller scales, or not end-to-end simulations.  Here we make projections for neutron star EOS constraints that can be expected from an O5 observing campaign with the LVK network in the A+ configuration, using the best current estimates of the BNS merger rate, population and EOS. We simulate a realistic population of BNS coalescences detectable in three years of continuous three-detector operation~\citep{LVKobsplan}, considering three possibilities for the assumed NS EOS, supporting relatively compact, typical and diffuse NSs, respectively. We calculate the source parameter uncertainties for each detected BNS signal in our simulated population using reduced-order-quadrature-- (ROQ--) accelerated GW parameter estimation (PE)~\citep{CanizaresField2015,SmithField2016,PhysRevD.104.063031}, as implemented in \bilby~\citep{AshtonHubner2019}. We then perform Bayesian inference of the EOS~\citep{LandryEssick2020} using \lwp \citep{lwp2024}, interpolating the PE samples for GW observation with a kernel density estimate and modeling the EOS phenomenologically as a Gaussian process~\citep{LandryEssick2019,EssickLandry2020}. 

Three years of LVK observation at A+ sensitivity typically results in 25 (within the range 0.9 to 450 estimated in Section \ref{sec:simulatedBNSpopulation}) BNS detections with a network signal-to-noise ratio (SNR) greater than 11.2. 
For a signal to qualify as a detection, we require not only a network SNR exceeding 11.2 but also that the SNR in each detector surpasses 5. This criterion ensures sufficient contributions from all detectors while maintaining high confidence in the detection significance. The network threshold itself is empirically calibrated, considering detector sensitivities, noise characteristics, and false alarm rates, as discussed in recent analyses \citep{Mould:2023eca}.
Analyzing these observations jointly in the \lwp, a hierarchical Bayesian inference framework, we find that the three equations of state can be marginally distinguished with necessary corrections. 
Despite the projected O5-era BNS observations thus promising incremental gains in our EOS knowledge, we conclude that precision constraints on the EOS from gravitational waves must wait until the proposed A\#~\citep{ObservingScenarios} or Voyager~\citep{AdhikariArai2020} upgrade to the LVK network, or until planned next-generation ground-based GW detectors like Einstein Telescope~\citep{MaggioreVanDenBroeck2020} and Cosmic Explorer~\citep{EvansAdhikari2021} come online.

Our study is organized as follows: Sec.~\ref{sec:methodology} describes our methodology; Sec.~\ref{sec:pe} presents the PE results for the simulated BNS population; and Sec.~\ref{sec:eosresults} presents the results of the EOS inference. We draw our conclusions and discuss them in Sec.~\ref{sec:discussion}.
\section{Methods}\label{sec:methodology}

\subsection{Simulated BNS Population}\label{sec:simulatedBNSpopulation}

Following~\citet{LandryRead2021,O3bPop}, we simulate a population of BNS coalescences by assuming a uniform NS mass distribution between $1,M_\odot$ and the Tolman-Oppenheimer-Volkoff (TOV) maximum mass, pairing them randomly into binaries. The NS spins are oriented isotropically and sampled uniformly in dimensionless magnitude up to $0.05$, consistent with the spins of known Galactic double NSs \citep{Ozel:2012ax, Kiziltan:2013oja} that will merge within a Hubble time. 

We distribute the sources uniformly over the sky and in comoving volume out to a distance of 460 Mpc, which is well beyond the BNS range of the LIGO-Virgo-KAGRA (LVK) network at O5 sensitivity~\citep{ObservingScenarios}. The chosen population model is summarized in Table~\ref{tab:injection_and_pe_priors}. The simulated signals are injected using \bilby into Gaussian noise representative of A+ sensitivity \citep{LVKcurvesAdv}. The used noise curves are \texttt{APlusDesign.txt} for LIGO Hanford and Livingstone, \texttt{avirgo\textunderscore O5high\textunderscore NEW.txt} for Virgo, and \texttt{kagra$\_$80MPc.txt} for KAGRA. 

Later in Sec. \ref{sec:eosresults}, we adopt an astrophysical BNS merger rate of  10 to 1700 $Gpc^{-3} yr^{-1}$~\citet{O3bPop} to constrain EOSs.  Translating this merger rate into the number of events, we expect to observe 0.9 to 150 and 2.5 to 450 BNS mergers for a BNS range of 240 Mpc and 345 Mpc, respectively, during a 3-year O5 observing run. Combined with the observations in O4, we assume around 25 BNSs will be observed and jointly analyze them to see how well the three different EOSs can be recovered.

\begin{figure}[t!]
\centering
\includegraphics[width=\linewidth]{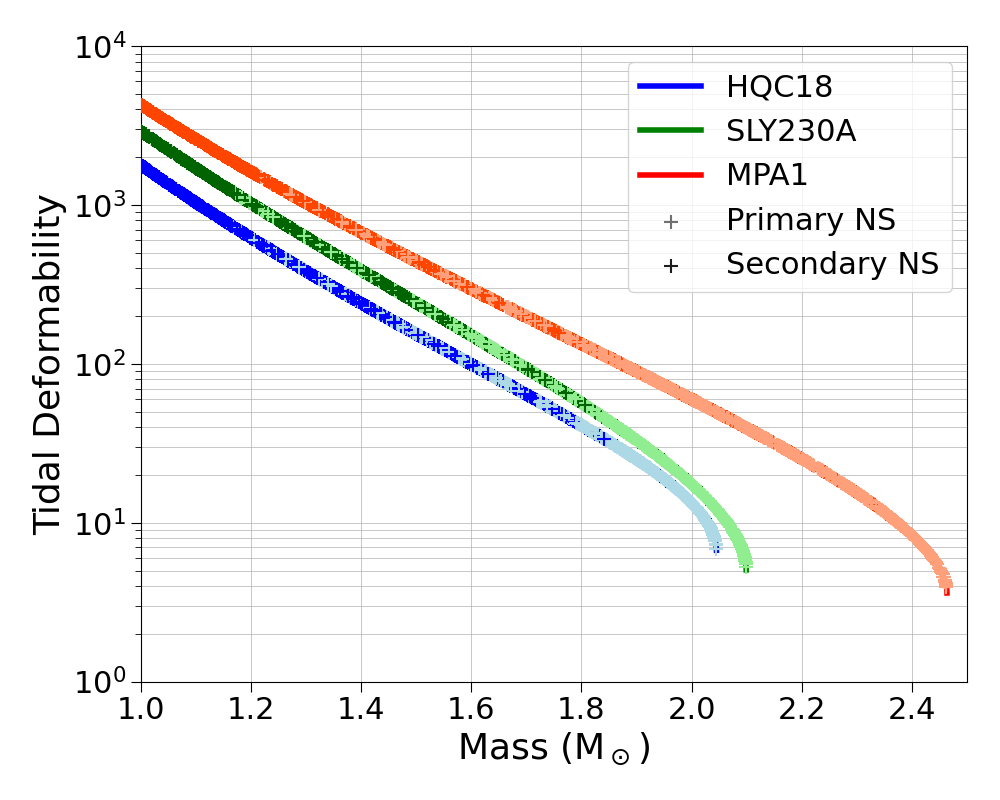}
\caption{Distribution of masses and tidal deformabilites for the components of the 3,000 BNS injections, with 1,000 injections per EOS. Light-colored crosses represent the primary NSs, while the dark-colored crosses denote the secondary NSs. The blue, green, and red colors correspond to the three EOS models.
\label{fig:injections_EOS}}
\end{figure}

The tidal deformability of each NS in the simulated population is determined by its mass and an assumed EOS. The EOS also sets the TOV mass, which bounds the NS mass spectrum from above. We consider three different choices for the EOS from among well-established candidate models in the nuclear theory literature: \textsc{hqc18}\citep{BaymFurusawa2019}, \textsc{sly230a}\citep{ReinhardFlocard1995}, and \textsc{mpa1}\citep{MutherPrakash1987}, which represent soft, average, and stiff EOSs, respectively. For each EOS, we solve the TOV equations\citep{Tolman1939,OppenheimerVolkoff1939} and the equation for quadrupolar tidal deformation~\citep{FlanaganHinderer2008,LandryPoisson2014} to compute the mass--$\Lambda$ sequence for stable, nonrotating NSs. We interpolate this sequence linearly and use the resulting $\Lambda(m)$ relation to assign the tidal deformability based on NS mass. The distributions of NS mass and tidal deformability resulting from this procedure are plotted in Figure~\ref{fig:injections_EOS} for each of the three EOSs considered.

For each of the three resulting populations, we generate multiple realizations of one year's worth of detectable BNS mergers. For every simulated NS binary, we inject a simulated gravitational waveform into Gaussian noise at A+ sensitivity for the LVK network, using the IMRPhenomPv2$\_$NRTidalv2 waveform model~\citep{DietrichSamajdar2019}.

Figure~\ref{fig:injections_EOS} shows the injected values of the tidal deformability ($\Lambda$) as a function of the masses for the two NSs in each BNS system. We have 3,000 injections in total, divided between the three equations of state (EOS): \textsc{hqc18}, \textsc{sly230a}, and \textsc{mpa1}, with 1,000 injections per EOS. Since each BNS consists of two NSs, there are 2,000 points per EOS, resulting in 6,000 points (crosses) on the plot. These injections cover the entire allowed EOS parameter space.

The EOS curves themselves, though not perfectly visible due to the number of crosses, show the hierarchy among the three EOSs considered. \textsc{hqc18} is the lowest, starting with a tidal deformability of about 2,000 for $1\,\rm{M}_{\odot}$ and dropping down to about 6 at $2\,\rm{M}_{\odot}$. The \textsc{mpa1} curve sits the highest, starting around 4,000 for $1\,\rm{M}_{\odot}$ and decreasing to about 4 at $2.5\,\rm{M}_{\odot}$. In between, \textsc{sly230a} begins around 3,000 for $1\,\rm{M}_{\odot}$ and ends near 5 at $2.1\,\rm{M}_{\odot}$. These curves exhibit a similar behavior across the mass range: a relatively flat slope in logarithmic scale before rapidly declining at the high-mass end.

Regarding the injection points, the primary NSs, which have higher masses, tend to be located in the lower tidal deformability regions (the far end of each EOS curve). Although it seems like the primary NSs cover a smaller part of the curves, this is because the secondary NSs, which are less massive, are plotted over them. Both primary and secondary NSs cover a significant range of each EOS curve. The secondary NSs mostly occupy the beginning of the curves, before the rapid decline in $\Lambda$. The steep final parts of the EOS curves are less densely populated as expected, since we are injecting uniformly in mass, not in $\Lambda$, which naturally leads to fewer points where the tidal deformability drops rapidly. 

This plot shows that the injections cover the allowed mass-tidal deformability parameter space for each EOS quite effectively, providing us a solid foundation for our analysis and subsequent statistics.

\subsection{GW Parameter Estimation}\label{sec:GW PE}

We applied the ROQ techniques \citep{CanizaresField2015,SmithField2016,PhysRevD.104.063031} embedded in \bilby to perform PE of the 3,000 GW injections, using the \texttt{IMRPhenomPv2$\_$NRTidalv2} model for recovery. The ROQ bases for this waveform model were taken from~\cite{PhysRevD.108.123040}. A simulated GW signal is considered to have a duration of 128-second if it falls within the chirp mass and frequency ranges for which the ROQs were constructed: chirp mass $\mathcal{M}{\rm c}\in[1.65,\ 2.60]\ M_{\odot}$,$ \ \text{lower frequency}\ f_{\text{min}}=20\ \text{Hz}$,$\ \text{and upper frequency}\ f_{\text{max}}=4,096\ \text{Hz}$. 
Equivalently, a signal is considered to have a duration of 256-second if $\mathcal{M}{\rm c}\in[0.98,\ 1.65]\ M_{\odot}$, with the same frequency range. 

For a confident detection of the GW signals, we assume a source is detected if single-detector SNR is larger than 4 in at least two detectors and the network SNR is larger than 11.2. This SNR threshold is consistent with the LVK Collaboration and their past detections~\citep{KAGRA:2013rdx,LIGOScientific:2018mvr,LIGOScientific:2020ibl,KAGRA:2021vkt}. 

Compared to the standard method in \bilby, the ROQ method accelerates likelihood calculations with 

\begin{widetext}
\begin{table*}[htb!]
\centering
\caption{\label{tab:injection_and_pe_priors}Distributions of simulated gravitational wave parameters and their parameter estimation priors}
\label{tb:distributions}
\begin{tabular}{l*{6}{l}l}
\hline\hline
Parameter (Symbol) [Unit] \quad  & Injection configuration \quad  & PE prior
\\ \hline
Source-frame primary NS ($m_1)\ [M_{\odot}]$ &  Uniform [1, ${m_{\rm TOV}}^{*}$] with $m_1\geq m_2$ & Determined by $\mathcal{M}_{\rm c}$ and $q$\\
Source-frame secondary NS ($m_2)\ [M_{\odot}]$ &  Uniform [1, $m_{\rm TOV}$]  & Determined by $\mathcal{M}_{\rm c}$ and $q$\\
Source-frame chirp mass  ($\mathcal{M}_{\rm c})\ [M_{\odot}]$ & 
$\frac{(m_1m_2)^{3/5}}{(m_1+m_2)^{1/5}}$ & \begin{tabular}{l} \hspace{-0.95cm} Uniform [1.6, 2.6] $M_{\odot}$ for 128 s \\ \hspace{-0.95cm} Uniform [0.98, 1.7] $M_{\odot}$ for 256 s\end{tabular}\\
Source-frame mass ratio ($q$) &  
$m_2/m_1$ & Uniform [0.125, 1] \\
Dimensionless primary NS spin  ($a_{1}$)  & Uniform [0, 0.05] & Uniform [0, 0.05] \\
Dimensionless secondary NS spin ($a_{2}$)  & Uniform [0, 0.05] & Uniform [0, 0.05] \\
Primary NS tilt ($\theta_1$) [radian] & \text{Uniform Sine} $[0,\pi]$ & Same as injection \\
Secondary NS tilt ($\theta_2$) [radian] & \text{Uniform Sine} $[0,\pi]$ & Same as injection \\ 
Relative spin azimuthal angle ($\phi_{jl}$) [radian] & \text{Uniform} $[0,2\pi]$ & Same as injection \\ 
Spin phase angle ($\phi_{12}$) [radian] & \text{Uniform} $[0, 2\pi]$ & Same as injection \\ 
Luminosity distance  ($d_L)\ [{\rm Mpc}]$ & Uniform in square [10, 460]\ \text{Mpc}  & \text{Square power law} [1, 1,000]\ Mpc \\
Right ascension  ($\alpha$) [radian] & Uniform [0, $2\pi$] & Uniform [0, $2\pi$] \\
Declination ($\delta$) [radian] & Uniform Cosine [$-\pi/2, \pi/2$] & Uniform Cosine \\
Inclination angle ($\theta_{\text{JN}}$) [radian] & Uniform Sine [0, $\pi$] & Uniform Sine [0, $\pi$]\\
Polarization  ($\Psi$) [radian] & Uniform [0,$\pi$] & Uniform [0, $\pi$] \\
Coalescence phase  ($\phi$) [radian] & Uniform [0, $2\pi$] & Marginalized \\
Geocenter time ($t_c$) [second] & Trigger time & \text{Uniform} [\text{trigger time} - 0.1, \text{trigger time} + 0.1]\\
Tidal deformability of primary NS ($\Lambda_1$) & Determined by $m_1$ and EOS & Uniform [0, 5,000] \\
Tidal deformability of secondary NS ($\Lambda_2$) & Determined by $m_2$ and EOS & Uniform [0, 5,000] \\
\hline\hline
\end{tabular}

\raggedright
\vspace{0.2cm}
* The TOV mass $m_{\rm TOV}$ depends on the EOS model. It is  $2.05\ M_{\odot}$ for \textsc{hqc18}, $2.10\ M_{\odot}$ for \textsc{sly230a}, and $2.47\ M_{\odot}$ for \textsc{mpa1}. The ranges for $\Lambda_1$ and $\Lambda_2$ are $[6.86, 795.23]$ and $[10.50, 1792.72]$ respectively for \textsc{hqc18}, $[5.25, 1395.15]$ and $[9.80, 2921.92]$ for \textsc{sly230a}, and $[3.97, 1744.57]$ and $[8.15, 4256.29]$ for \textsc{mpa1}.
\end{table*}
\end{widetext}
\texttt{IMRPhenomPv2$\_$NRTidalv2} model by a factor of 230-550, which are the most time-consuming part of a PE run, reducing the analysis time for a single GW event from several months to a few hours for a 128-second signal and about 1 day for a 256-second signal. The parameter estimation of each set of 1,000 events (mostly 256-second) was completed within 6 days on \texttt{Cedar} using two Intel Xeon CPU E5-2683 v4 Broadwell @ 2.1GHz chip and 32 cores each, considering the overall time including task queuing. Each PE run requests 1 core and 8 GB RAM.

The parameter distributions of the injected GWs and their PE priors are detailed in Table~\ref{tab:injection_and_pe_priors}.

\subsection{EOS Inference}

Given the PE posterior samples 
in component masses and tidal deformabilities for each detected BNS event in our simulated population, we perform a hierarchical Bayesian inference for the common NS EOS following~\citet{LandryEssick2020}. The posterior probability of an EOS proposal $\varepsilon$, jointly conditioned on all of the GW observations $d = \{ d_i \}$ for $i=1,...,N$, is

\begin{equation} \label{posterior}
    P(\varepsilon | d) \propto \pi(\varepsilon) \prod_i \mathcal{L}(d_i | \varepsilon)
\end{equation}
with the EOS likelihood
\begin{equation} \label{likelihood}
\begin{split}
    \mathcal{L}(d_i | \varepsilon) =& \int \mathcal{L}(d_i | m_{1,2}^i,\Lambda_{1,2}^i) \pi(m_{1,2}^i,\Lambda_{1,2}^i | \varepsilon, \lambda) \pi(\lambda) \\ 
    &\times \zeta^{-1}(\lambda) dm_{1,2}^i d\Lambda_{1,2}^i d\lambda, 
\end{split}
\end{equation}
where $\mathcal{L}(d_i | m_{1,2}^i,\Lambda_{1,2}^i)$ is the GW PE likelihood for the $i$th event. The distributions $\pi(\varepsilon)$, $\pi(\lambda)$, $\pi(m_{1,2}^i,\Lambda_{1,2}^i | \varepsilon, \lambda)$ are the priors on the EOS, population model, and $i$th-event GW parameters, respectively. The quantity

\begin{equation}
    \zeta(\lambda) = \int P_{\rm det}(m_{1,2},\Lambda_{1,2}) \pi(m_{1,2},\Lambda_{1,2}|\varepsilon,\lambda) dm_{1,2} d\Lambda_{1,2} ,
\end{equation}
the fraction of the population that is detectable, accounts for selection effects. The probability of detecting an event, $P_{\rm det}(m_{1,2},\Lambda_{1,2}) \approx P_{\rm det}(m_{1,2})$, is simply a step function in SNR---which is almost entirely determined by the chirp mass---and is thus virtually independent of $\Lambda_{1,2}$.

In our implementation of the inference, we model the EOS phenomenologically as a Gaussian process, following~\citet{LandryEssick2019,EssickLandry2020}. Specifically, we used the Gaussian process EOS samples labeled ``PSR'' made available in the data release for~\citet{LegredChatziioannou2021}. These are conditioned on the mass measurements for the two heaviest known Galactic pulsars, PSR J0740+6620~\citep{FonsecaCromartie2021} and PSR J0348+0432~\citep{AntoniadisFreire2013}, which enforce $M_{\rm TOV} \gtrsim 2\,M_\odot$. They serve as samples from $\pi(\varepsilon)$.

For each EOS sample, we compute the likelihood in Eq.~\ref{likelihood} for each GW event as a Monte Carlo sum using the software package \lwp. We fix the population model $\lambda$ in the inference to match the one used to create the simulated BNS population, so that $\pi(\lambda)$ is a Dirac delta function; this ensures that the EOS inference is unbiased, despite not simultaneously inferring the NS mass distribution (cf.~\citet{WysockiO'Shaughnessy2020}). Because the selection function $P_{\rm det}$ is approximately independent of tidal deformability
, the EOS inference is not impacted by GW selection effects when $\lambda$ is fixed, such that $\zeta$ is the same for all $\varepsilon$ and can be ignored. The posterior probability of each EOS sample then follows from the product in Eq.~\ref{posterior}.

Concerning the configuration of the EOS inference, we set the number of marginalization points to 100. For each GW event, we limited the maximum number of posterior samples to 5,000. The EOS samples were read from the file \texttt{LCEHL$\_$EOS$\_$posterior$\_$samples$\_$PSR.h5} (\cite{LegredChatziioannou2021}), and we processed all 10,000 samples present in it in our inference. We ensured that the effective sample size $N_{\rm{eff}}$, which quantifies the number of unique EOS samples that significantly contribute to the likelihood calculation, was above 100 for each combined event group to maintain the robustness of our results. The effective sample size is calculated using the formula
\begin{equation} 
N_{\rm{eff}} = \frac{\left(\sum_i w_i\right)^2}{\sum_i w_i^2}, 
\end{equation}
where the weights $w_i$ are the marginalized likelihoods computed by \lwp for each EOS sample $i$.

For EOS inference, we processed all 1,000 events for each of the three EOS models (\textsc{hqc18}, \textsc{sly230a}, and \textsc{mpa1}) in three nearly equal batches. The computations were performed in 2024 using the CEDAR computing cluster. The runs for \textsc{hqc18} were conducted from May 15th to 21st, for \textsc{sly230a} from May 21st to 27th, and for \textsc{mpa1} from May 18th to 24th, all in 2024. Each batch took approximately two days to complete, with around 300 events processed concurrently. If run individually, the \lwp analysis for a single event took approximately 20 hours to 3 days. This parallel processing significantly reduced the overall computation time, enabling us to efficiently handle the large number of events.
\\

\section{PE results}\label{sec:pe}

\begin{figure*}[htp]
  \centering
  \subfigure{\includegraphics[width=\textwidth]{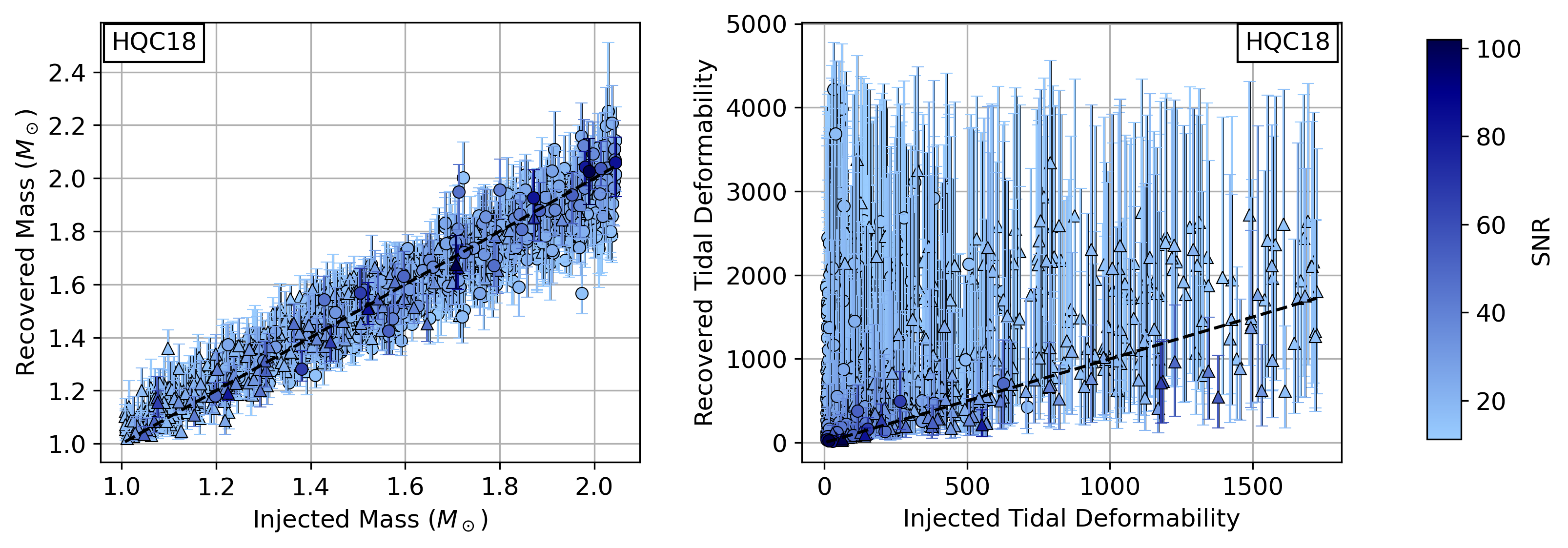}}
  \subfigure{\includegraphics[width=\textwidth]{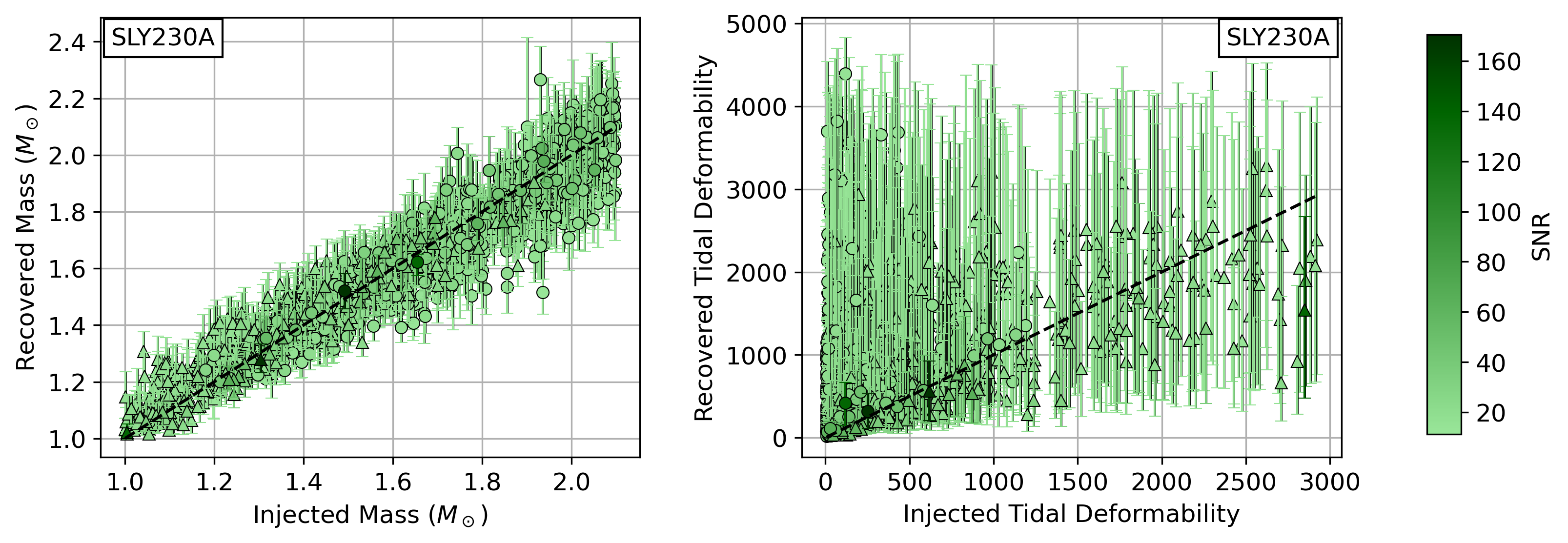}}
  \subfigure{\includegraphics[width=\textwidth]{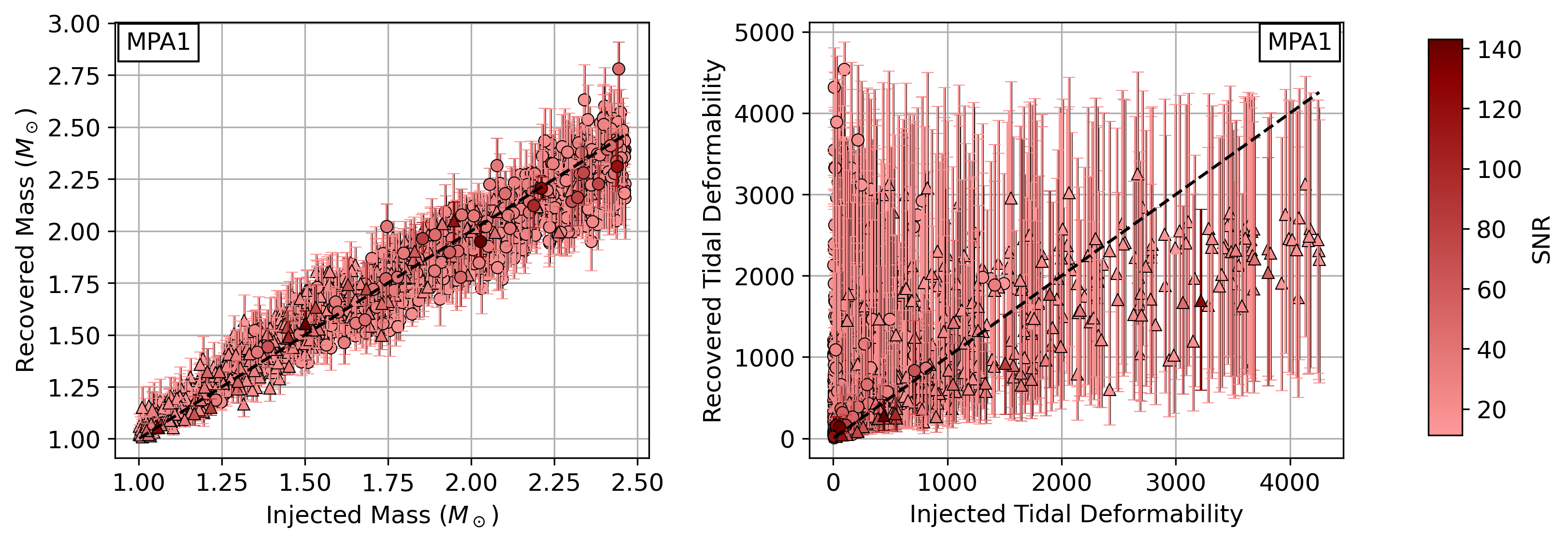}}
  \caption{Comparison of injected versus recovered component masses (left column) and tidal deformabilities $\Lambda$ (right column) for the three EOS models, considering only events with SNR$>$11.2. The x-axis represents the injected values, and the y-axis shows the posterior medians with error bars. Both primary (circles) and secondary (triangles) NSs are included. The dashed line denotes perfect recovery (injected equals recovered). Highest combined SNR: 102 (\textsc{hqc18}), 170 (\textsc{sly230a}) and 143 (\textsc{mpa1}).
\label{fig:injected_vs_posterior_snr_11.2}} 
\end{figure*}

In our analysis, we considered only events with a SNR greater than 11.2, consistent with our detection criteria (see Section~\ref{sec:GW PE}). This selection resulted in 617, 635, and 676 events for the \textsc{hqc18}, \textsc{sly230a}, and \textsc{mpa1} EOS models, respectively. These high-SNR events provided robust data for constraining the EOS using the recovered mass and tidal parameters from the GW injections.

Figure~\ref{fig:injected_vs_posterior_snr_11.2} presents a comparison between the injected parameters and the posterior estimates for our simulated GW events. The figure consists of six panels arranged in three rows, each corresponding to one of the NS EOS models: \textsc{hqc18} in blue, \textsc{sly230a} in green, and \textsc{mpa1} in red. In each row, the left panel displays the comparison between the injected and posterior component masses of the two NSs, while the right panel shows the comparison for the component tidal deformabilities $\Lambda$.

All plots include a dashed line representing perfect recovery, serving as a reference. The data points are color-coded according to the combined SNR of the injections, ranging from light color (lower SNR) to dark color (higher SNR). 

In the mass recovery plots (left column), we observe that the injected and posterior masses align closely along the equality line across all EOS models, indicating quite accurate mass estimation. The error bars are relatively small, suggesting high precision in the mass measurements. While there is a slight variation in the mass range covered by each EOS—\textsc{mpa1} and \textsc{sly230a} span a wider mass range than \textsc{hqc18}—this does not significantly impact the overall accuracy of the mass recovery.

In contrast, with EOS-agnostic PE, we recovered the uninformative tidal priors in the vast majority of the cases. The tidal deformability plots (right column) exhibit a larger spread and reveal noticeable differences between the EOS models. The posterior estimates of $\Lambda$ show significant uncertainties, with error bars considerably larger than those for the masses. Despite these uncertainties, most data points remain compatible with the equality line within their error margins. However, we observe a systematic trend where the tidal deformability tends to be overestimated when the injected value is small and underestimated when the injected value is large, particularly at lower mass. This behavior is more noticeable for the \textsc{sly230a} and \textsc{mpa1} EOS models, which predict larger tidal deformability compared to \textsc{hqc18}.

These systematic biases in the estimation of tidal deformability could potentially impact the combined EOS inference from multiple events. If not properly accounted for, they may introduce biases in the inferred $\Lambda$-mass relationship, affecting the constraints on the NS EOS. Therefore, it is crucial to carefully handle the tidal deformability posteriors in our hierarchical Bayesian analysis to mitigate these biases.
\section{EOS Constraints}\label{sec:eosresults}





Before proceeding with the detailed analysis, it is essential to discuss the role of the effective sample size ($N_{\text{eff}}$) in our EOS inference using the \lwp software. The parameter $N_{\text{eff}}$ quantifies the number of unique EOS samples from our prior distribution (comprising 10,000 EOS samples) that significantly contribute to the likelihood calculation. To ensure robust and reliable EOS constraints, we set a threshold of $N_{\text{eff}}>100$ when combining events. This choice guarantees that a sufficient diversity of EOSs influences the inference, preventing the results from being dominated by a small subset of EOS samples, which could lead to biased or incorrect conclusions.
When attempting to combine all available events, especially those with lower SNR, the value of $N_{\text{eff}}$ can decrease significantly. A low $N_{\text{eff}}$ indicates that only a few EOS samples are compatible with the data from all events, increasing the risk of overfitting and erroneous EOS recovery.
Thus, this requirement impacts the number of groups we can consider when increasing the number of events per group, as seen in some of the plots in this section (e.g., Figure~\ref{fig:eoswithnumberofevents_allevents}). For instance, when forming groups of 30 events, the number of groups decreases not only because of the limited total number of events available but also due to the $N_{\text{eff}}$ threshold. This effect is particularly noticeable for the \textsc{mpa1} EOS model. Despite \textsc{mpa1} having, on average, higher SNRs and more events that pass the initial SNR threshold, we end up with only 4 groups when considering groups of 30 events. This is because most of the potential groups do not satisfy the $N_{\text{eff}}>100$ condition when combining the posteriors of the 30 events. Consequently, we do not include these groups in our final results to maintain the robustness of the EOS inference.

In the analysis presented below, all the results correspond to averaging over all groups of a given number of events (e.g., 10, 20, or 30 events) for each EOS model, using the medians and 90\% credible intervals from the combined posteriors of the events that satisfy both the $N_{\text{eff}}>100$ and SNR$>11.2$ criteria. This approach ensures that the statistical conclusions drawn are based on sufficiently representative samples, enhancing the reliability of our EOS constraints.

\subsection{SNR's Effect}

We found that SNR determines how well we can measure EOS slopes. When SNRs are lower, even with 600 events, we have bias on every EOS model’s slope. When SNR$>$35, we can measure slopes very accurately. There is not much difference between SNR$>$25 and SNR$>$35 although the numbers of events are almost tripled. 

\subsection{Effect of the Number of Events}

\begin{figure*}[ht!]
\centering
\includegraphics[width=\textwidth]{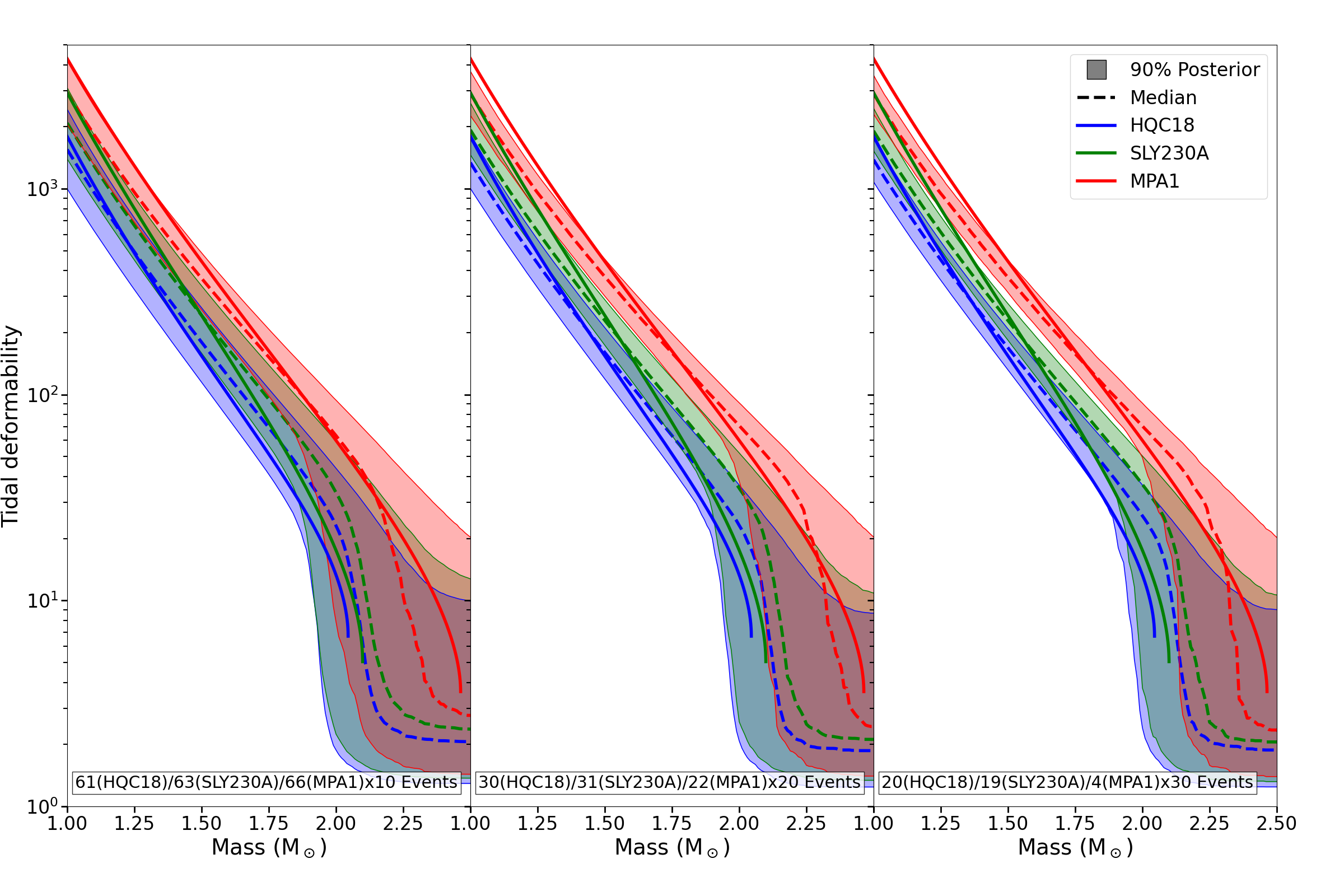}
\caption{Combined posterior constraints on the NS EOSs, averaged over groups of 10 (left panel), 20 (middle), and 30 (right) events with $\text{SNR}>11.2$, considering only groups where $N_{\text{eff}}>100$. The numbers before the parentheses indicate the number of groups used for each EOS model. The shaded regions repreent the 90\% credible intervals, and the dashed lines show the median recovered EOSs. The solid lines correspond to the injected EOS curves.
\label{fig:eoswithnumberofevents_allevents}}
\end{figure*}

Figure~\ref{fig:eoswithnumberofevents_allevents} illustrates the effect of varying the number of events included in the combined posterior analysis on the recovery of the NS EOS. The figure is divided into three panels, corresponding to groups of 10, 20, and 30 events from left to right, respectively. For each EOS model—\textsc{hqc18} (blue), \textsc{sly230a} (green), and \textsc{mpa1} (red)—we selected randomized groups of events that satisfy the conditions of a network SNR greater than 11.2 and an effective sample size $N_{\rm eff} > 100$. Due to these stringent criteria, the number of groups considered for each EOS and group size is smaller than the total number of simulated events, and this number is indicated at the bottom of each panel.

In each panel, the shaded regions represent the 90\% credible intervals derived from the combined posterior distributions for each EOS model. The dashed lines correspond to the median recovered EOS for the respective models, matching the color coding. The injected EOS curves are indicated by solid lines.

For groups consisting of 10 events, we observe that the recovery of the EOS is relatively good across all models. The 90\% credible intervals encompass the true EOS values, especially for \textsc{hqc18} and \textsc{sly230a}. However, for \textsc{mpa1}, there is a slight underestimation of the tidal deformability at lower masses within the 90\% credible regions. Notably, the median recovered EOS curves exhibit systematic biases: they are lower than the injected EOS at low masses and higher at high masses for \textsc{hqc18} and \textsc{sly230a}, whereas for \textsc{mpa1} the median consistently underestimates the true EOS across the mass range.

Increasing the group size to 20 events, the medians of the recovered EOS remain relatively stable, maintaining the biases observed with 10 events. This stability suggests that the systematic biases are inherent to the PE process and are not significantly reduced by simply increasing the number of events. However, the 90\% credible intervals become narrower, tightening around the biased medians. This indicates that while our confidence in the median estimates increases with more events, the underlying biases is still present. At low masses, only the \textsc{hqc18} EOS is accurately recovered, whereas the other EOS models continue to exhibit underestimation. At higher masses, all EOS models are better recovered, with the injected EOS lying within the 90\% credible regions, although the biases in the medians remain.

When the group size is further increased to 30 events, the results show only marginal changes compared to the 20-event groups. This plateau in improvement suggests that we have reached the limit of constraint achievable with the used detector sensitivities (A+ sensitivity curves) and the number of events considered. To obtain more stringent constraints on the EOS, we would require either detectors with better sensitivity, louder events or a substantially larger number of events—on the order of several hundred—which is not feasible in the A+ era.

\subsection{Effect of Event Selections}

\begin{figure*}[ht!]
\centering
\includegraphics[width=\textwidth]{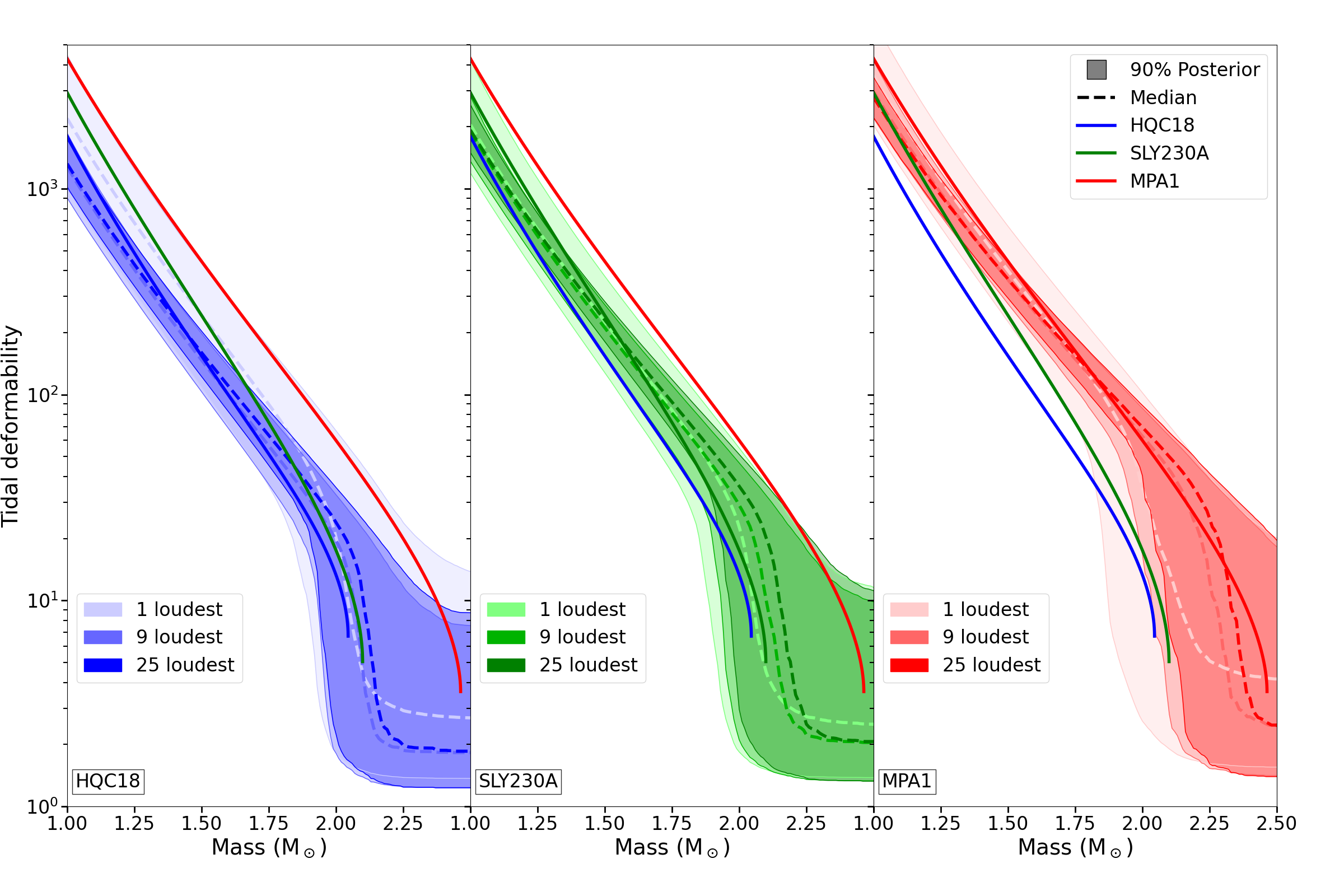}
\caption{Constraints on the three EOS models by averaging over the 1, 9 and 25 loudest events within groups of 25 events, all with $\text{SNR}>11.2$ and $N_{\text{eff}}>100$. The shaded regions indicate the 90\% credible intervals, with darker shades representing results that include more loudest events. The dashed lines are the median recovered EOSs, and the solid lines depict the injected EOS curves. 
\label{fig:combined_plot_N_loudests}}
\end{figure*}

Figure~\ref{fig:combined_plot_N_loudests} shows how the selection of the loudest GW events influences the recovery of the EOS when considering groups of 25 events—representative of approximately three years of observations in the A+ era. Instead of including all events in each group, we focus on the $N$ loudest events based on their SNR, aiming to assess how the number of high-quality detections affects the EOS constraints. This approach provides insight into the evolution of our EOS understanding as more significant detections are accumulated.

The figure is divided into three panels, each corresponding to one of the EOS models: \textsc{hqc18} (left), \textsc{sly230a} (middle), and \textsc{mpa1} (right). In each panel, we present the averaged results for the 1, 9, and 25 loudest events within the groups of 25, depicted from the lightest to the darkest shade of the respective EOS color.

For \textsc{hqc18}, we observe that increasing the number of loudest events included in the analysis leads to a decrease in the median recovered tidal deformability at low masses. This introduces a bias where the median underestimates the true EOS values more prominently as more events are considered. Conversely, at higher masses, including more events results in an increased median tidal deformability, causing an overestimation relative to the injected EOS.

In the case of \textsc{sly230a}, the median recovered EOS remains relatively constant at low masses regardless of the number of events included. However, at higher masses, the median increases significantly when more events are considered, leading to a substantial overestimation of the tidal deformability. The single loudest event case demonstrates less bias at high masses compared to the cases with 9 and 25 events.

For \textsc{mpa1}, the median recovered EOS consistently underestimates the injected EOS across the entire mass range when more events are included. Including additional loudest events slightly reduces the bias at higher masses, bringing the median closer to the true EOS. Nevertheless, at lower masses, we underestimate the EOS regardless of the number of events considered.

Across all three EOS models, we notice that the 90\% credible intervals become narrower as more events are included, while the systematic biases in the medians remains consistent. The differences between the results for the 9 and 25 loudest events are minimal, particularly at low masses, indicating that the constraints converge quickly with the inclusion of a relatively small number of high-SNR events. This suggests that under A+ detector sensitivity conditions, adding more than approximately 10 high-SNR events does not significantly enhance the EOS recovery.


\subsection{Effect of Chirp Mass and EOS Softness}

\begin{figure*}[ht!]
\centering
\includegraphics[width=0.95\textwidth]{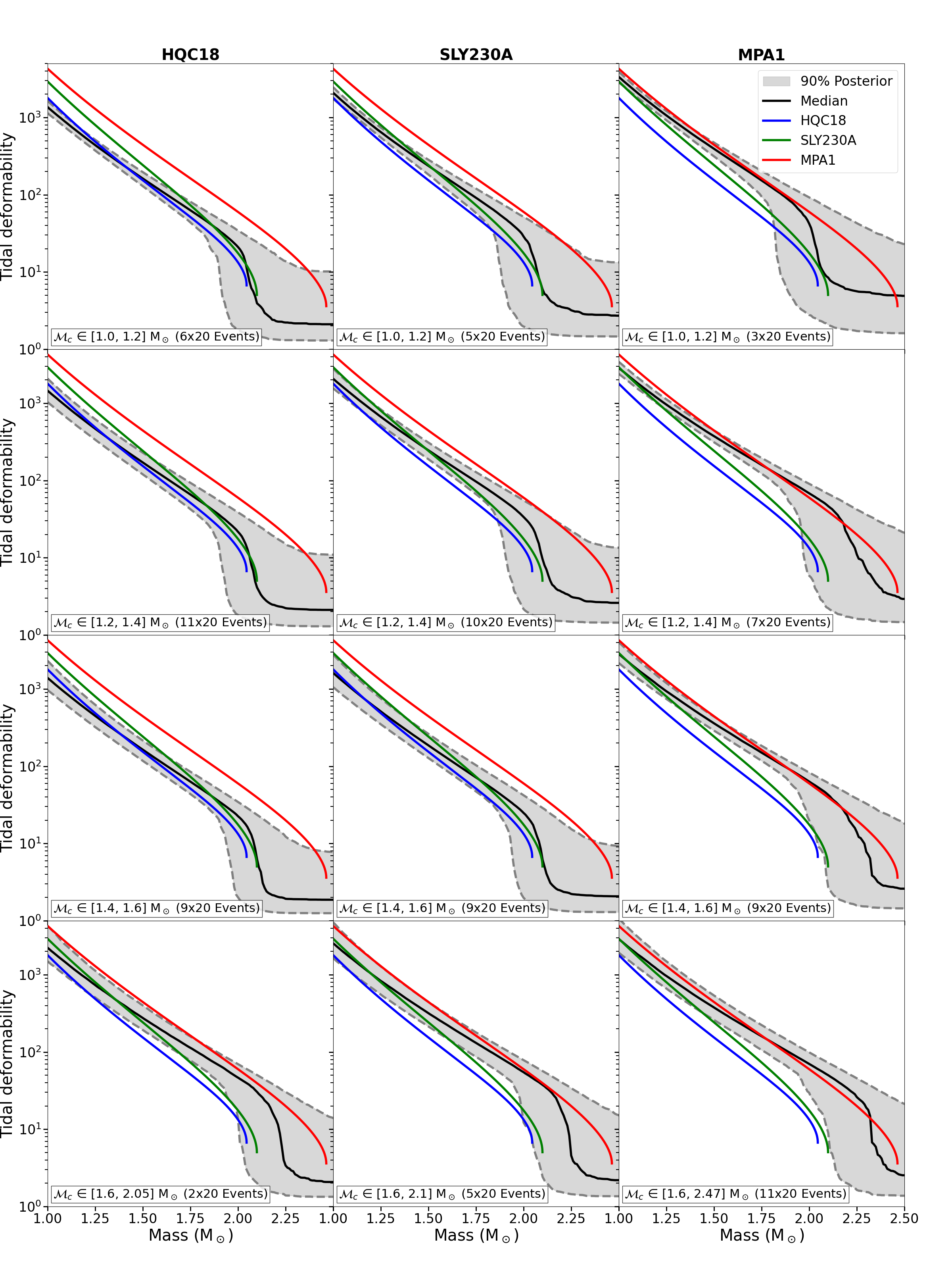}
\caption{Combined posterior EOS constraints averaged over groups of 20 events, with each geoup satisfying SNR$>11.2$ and $N_{\text{eff}}>100$. Each column corresponds to one of the EOS models, and each row represents a different range of injected $\mathcal{M}_{\rm c}$ ($[1.0, 1.2]\, {\rm M}_{\odot}$, $[1.2, 1.4]\, {\rm M}_{\odot}$, $[1.4, 1.6]\, M_{\odot}$, and $[1.6, \mathcal{M}_{\rm c}^{\rm max}]\, {\rm M}_{\odot}$, where $\mathcal{M}_{\rm c}^{\rm max}$ depends on the EOS model). The numbers in parentheses indicate the number of groups considered for each case. The gray shaded regions show the 90\% credible intervals, and the black lines are the median recovered EOSs.
\label{fig:eoswithchirpmassbins}}
\end{figure*}

\begin{figure*}[ht!]
\centering
\includegraphics[width=\textwidth]{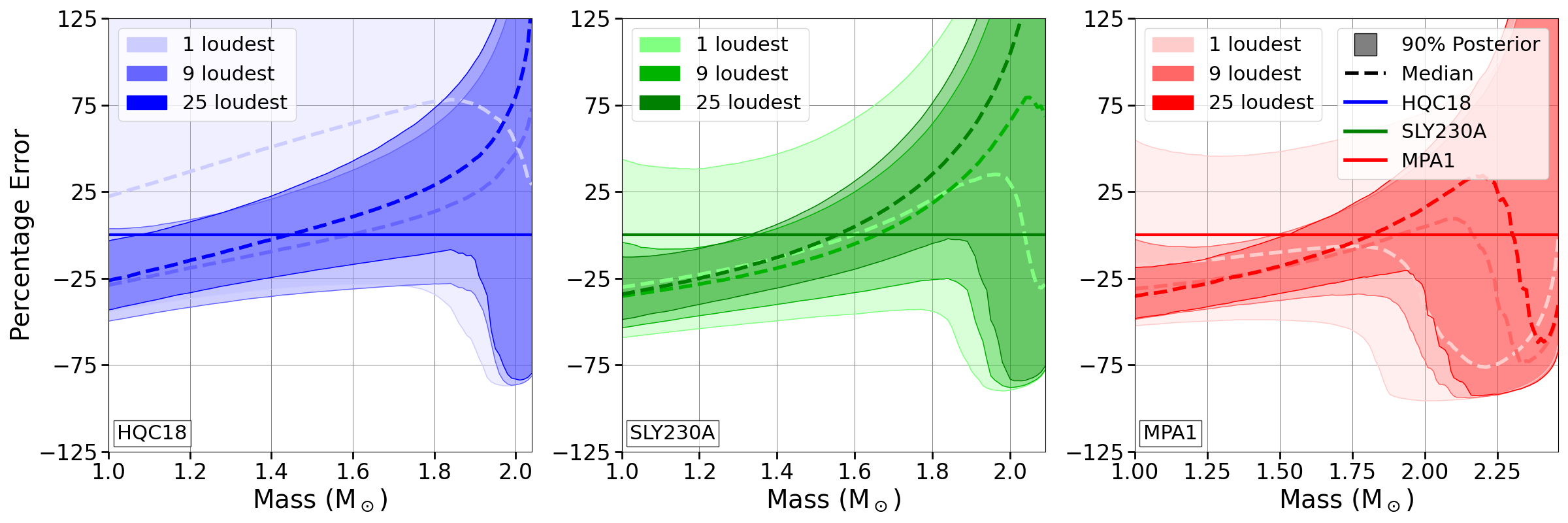}
\caption{Relative percentage error between the recovered and the injected tidal deformability values as a function of mass for each model, corresponding to Figure~\ref{fig:combined_plot_N_loudests}. 
\label{fig:combined_plot_N_loudests_corrections}}
\end{figure*}

Figure~\ref{fig:eoswithchirpmassbins} displays the combined posterior distributions for the three EOS models across four chirp mass intervals: $[1.0, 1.2]\, {\rm M}_{\odot}$ (first row), $[1.2, 1.4]\, {\rm M}_{\odot}$ (second row), $[1.4, 1.6]\, M_{\odot}$ (third row), and $[1.6, \mathcal{M}_{\rm c}^{\rm max}]\, {\rm M}_{\odot}$ (fourth row), where $\mathcal{M}_{\rm c}^{\rm max}$ denotes the upper chirp mass limit specific to each EOS model. In each panel, we present the combined posteriors from randomized groups of 20 events, selected after filtering events with $N_{\rm eff}>100$ and SNR$>11.2$. This approach emulates a realistic scenario for the upcoming O5 observing run, for a survey of 3 years and around 8 detections per year. 

The solid colored lines represent the true EOS curves for each model. The gray shaded regions indicate the 90\% credible intervals derived from the combined posterior distributions, while the black lines denote the median recovered EOS.

Examining the results, we observe that for both \textsc{hqc18} (left column) and \textsc{sly230a} (middle column), the two intermediate chirp mass bins (second and third rows) recover the injected curves reasonably well, with the distributions overlapping the real EOS within their 90\% credible intervals, albeit still displaying some bias. In contrast, the first panel (lowest chirp mass bin) exhibits a more noticeable offset at smaller masses, which is expected given that restricting the chirp mass range to small values tightly constrains lower masses and thus narrows the posterior there. The largest discrepancy occurs in the last panel (highest chirp mass bin), where the recovered EOSs deviate significantly from the injected curves over most of the mass range. They only converge back at the low- and high-mass ends. This suggests that biases become more obvious for the highest chirp masses, particularly near the maximum allowable mass for these softer EOS models.

Turning to \textsc{mpa1} EOS (right column), the overall trend appears better recovered, especially at the low- and high-chirp-mass bins. While this does not necessarily mean the 90\% credible intervals fully capture the injected curve, the recovered slope aligns more closely with the injected EOS across the entire mass range. In the last panel, the stronger agreement may stem from the broader mass range of the \textsc{mpa1} EOS, allowing for a larger range of chirp masses to be included in this bin and thus reducing the bias observed for the other two models at high chirp masses.



For each EOS, the tidal deformability is underestimated at lower NS mass and overestimated at higher mass. This could introduce systematic errors which need to be properly accounted for in the analysis.

\subsection{Analysis of Systematic Biases}

Figure~\ref{fig:combined_plot_N_loudests_corrections} quantifies the relative percentage error between the recovered tidal deformabilities and the injected tidal deformabilities for each model as a function of mass and as we increase the number of loudest events considered in groups of 25, as for Figure~\ref{fig:combined_plot_N_loudests}, providing a clearer understanding of how the systematic biases affect our estimations. This figure maintains the same panel arrangement and color coding as Figure~\ref{fig:combined_plot_N_loudests}.

For \textsc{hqc18}, the recovered tidal deformabilities generally remains consistent with the injected one within the 90\% credible intervals, regardless of the number of loudest events considered. A systematic bias exists: the tidal deformability is underestimated by roughly 25\% at low masses and overestimated by up to 125\% at high masses when considering the 25 loudest events case. The magnitude of this bias varies with the number of events. Although minimal changes appear between the 9 and 25 loudest events cases, more significant discrepancies arise when comparing with the single loudest event scenario.

In the \textsc{sly230a} case, the bias is slightly more significant. As more events are included, there is a marked underestimation of the tidal deformability at lower masses—on the order of 30\% below $1.55\,{\rm M}_\odot$ for the 25 loudest events case. At higher masses, the bias transitions to an overestimation, reaching about 125\%.

In \textsc{mpa1}, the median recovered tidal deformability consistently underestimates the injected one below $1.55\,{\rm M}_\odot$, irrespective of the number of events. This underestimation reaches about 30\% at low masses for the 25 loudest events set. Unlike in the \textsc{hqc18} and \textsc{sly230a} cases, however, the overestimation at higher masses declines after peaking at around 30\% near $2.15\,{\rm M}_\odot$, eventually giving way to an underestimation that can reach 60\%. Notably, the 9 loudest events set yields a similar trend, though the peak error is slightly shifted and reduced. By contrast, the single loudest event configuration shows consistent underestimation across the entire mass range. Interestingly, the percentage error curve for \textsc{mpa1} with 25 loudest events resembles the percentage error for \textsc{sly230a} in the single loudest event case. Meanwhile, in \textsc{sly230a}, this slope decreases as the number of loudest events increases, transitioning from a strong overestimation-to-underestimation crossover for fewer events to a gentler slope for 9 and then 25 loudest events. Such observations suggest that adding even more loudest events for \textsc{mpa1} might eventually yield a slope comparable to that seen for the other EOSs at 25 events, although we could not verify this due to limitations in $N_{\rm eff}$ (which we wanted larger than 100).

Overall, when examining the groups of 25 loudest events, all three EOSs exhibit a systematic shift from underestimating to overestimating the tidal deformability at some mass threshold. The location of this transition depends on the steepness of the EOS: steeper EOSs shift this crossover to higher masses. A distinguishing feature of \textsc{mpa1} lies in the subsequent reduction of the overestimation error after its peak, whereas the overestimation for \textsc{hqc18} and \textsc{sly230a} continues to grow toward their respective maximum allowed masses. These systematic biases—underestimation at low masses and overestimation at high masses—remain clearly visible in all cases. Moreover, increasing the number of high-SNR events does not necessarily mitigate the bias. However, it appears to converge toward a particular slope. This behavior suggests that factors beyond pure statistical uncertainties are driving these biases, including possible waveform modeling limitations, intrinsic EOS features, sensitivity-curve assumptions, or other systematic aspects of the data analysis pipeline.




\section{Discussion}\label{sec:discussion}

In this study, we investigated the potential of GW observations during the upcoming A+ era of the LVK network to constrain the NS EOS. By simulating 3,000 BNS mergers using three representative EOS models—\textsc{hqc18}, \textsc{sly230a}, and \textsc{mpa1}—we aimed to understand how well the EOS can be recovered from GW data and what factors influence the accuracy and precision of the constraints.
Our methodology was generating a realistic population of BNS systems with masses uniformly distributed between $1\,M_\odot$ and the maximum mass allowed by each EOS. The tidal deformabilities of the NSs were determined by their masses and the chosen EOS, ensuring that our simulations accurately reflected the physical properties of NSs as predicted by nuclear physics. GW signals from these BNS mergers were injected into simulated noise representative of the expected sensitivity during the A+ observing run, and we performed full PE using the ROQ techniques.

One of the findings from our PE analysis is that while NS mass is accurately recovered with high precision across all EOS models, tidal deformability exhibits significant uncertainties and systematic biases. Specifically, these biases show an underestimation of the tidal deformability at lower mass and an overestimation at higher masses, with the exact transition mass depending on the steepness of the EOS.
These systematic biases in the estimation of tidal deformability have important implications for EOS inference. When combining multiple events to constrain the EOS through hierarchical Bayesian inference, the biases can lead to systematic errors in the inferred mass–$\Lambda$ relationship. Our analysis showed that including more events in the analysis generally improves the precision of the EOS constraints, as evidenced by narrower credible intervals. However, the systematic biases in the median recovered EOS is present regardless of the number of events included. This suggests that simply increasing the sample size is insufficient to eliminate these biases.

We explored the effects of various factors on the EOS constraints, listed here:

\noindent 1. \textit{Effect of the number of events:} We analyzed groups of 10, 20, 30 events and found that while increasing the number of events tightens the credible intervals, the systematic biases in the recovered EOS medians remain. This indicates that statistical uncertainties are not the sole contributors to the biases and that systematic errors are influencing the results. The median EOS curves showed consistent patterns of underestimation at low masses and overestimation at high masses for certain EOS models, regardless of the sample size (see Figures~\ref{fig:eoswithnumberofevents_allevents}).

\noindent 2. \textit{Effect of event selection:} By focusing on the loudest events (those with the highest SNR), we observed that the inclusion of more high-quality detections does not significantly improve the EOS recovery beyond a certain point. The constraints converge quickly with the inclusion of approximately 10 loudest events, and adding more events brings minor effects. This suggests that under current detector sensitivity conditions, prioritizing the highest-SNR events is beneficial but has limited impact on mitigating systematic biases (see Figure~\ref{fig:combined_plot_N_loudests}~and~\ref{fig:combined_plot_N_loudests_corrections}).

\noindent 3. \textit{Effect of chirp mass and EOS softness:} We examined how the recovery of the EOS depends on the chirp mass of BNS systems and the softness of the EOS. For both \textsc{hqc18} and \textsc{sly230a}, the injected curves are reasonably well recovered at intermediate chirp mass (1.2 to 1.6 ${\rm M}_{\odot}$), despite noticeable biases. However, the lowest chirp masses exhibits a more obvious offset at smaller masses, while the highest chirp masses reveals significant deviations over much of the mass range. By contrast, for the \textsc{mpa1} EOS, the steepest one, the overall slope aligns more closely with the injected curve, particularly at low and high chirp masses, likely due to its broader maximum mass range. In all cases,  tidal deformability tends to be underestimated at lower NS mass and overestimated at higher mass, making it important to account for systematic effects when interpreting these results (see Figure~\ref{fig:eoswithchirpmassbins}).

The implications of our study are significant for the fields of GW astrophysics and nuclear physics. While the A+ era promises advancements in our understanding of NS physics, our results indicate that fully resolving the properties of NS matter will remain challenging. Nonetheless, a key highlight of our research is the potential to use statistical analysis from large simulations of BNS mergers with known injected parameter values and the given EOS models to refine and correct estimates toward the true EOS, as shown in Figure~\ref{fig:combined_plot_N_loudests_corrections}. In the A+ era, such as during O5, this method can enhance the accuracy of EOS constraints, given that the ROQ techniques have proven capable of handling the required large simulations in a desirable timescale—within a week.

In addition, achieving more accurate and unbiased EOS constraints will likely require: 

\noindent 1. \textit{Improved detector sensitivity:} Future observatories with significantly enhanced sensitivities, such as Cosmic Explorer and Einstein Telescope, will detect gravitational waves of higher SNRs and from a larger number of BNS mergers. These increased sensitivities will enable more precise measurements of tidal deformability and help mitigate some of the systematic biases observed. 

\noindent 2. \textit{Refined waveform models:} Developing more accurate waveform models that incorporate higher-order effects, such as tidal resonances, higher multipole moments, and improved descriptions of matter effects, will enhance the fidelity of PE and may reduce systematic errors. 

\noindent 3. \textit{Advanced analysis techniques:} Employing novel statistical methods, such as machine learning algorithms or improved hierarchical Bayesian inference frameworks, may help to better account for systematic uncertainties and extract more reliable constraints on the EOS.

\noindent 4. \textit{Multimessenger observations:} Combining GW data with electromagnetic observations of BNS mergers, such as kilonovae and short gamma-ray bursts, can provide complementary constraints on the EOS and help break degeneracies inherent in GW measurements alone.

It is noteworthy that the ROQ techniques used in our analysis significantly reduced the computational cost of PE, making it feasible to process thousands of simulated events. This computational efficiency is essential for handling the expected large number of GW detections, particularly low-mass systems like subsolar-mass BBHs and BNSs, in the upcoming observing runs.

In conclusion, our study demonstrates the potential and limitations of constraining the NS EOS using GW observations during the A+ era. While systematic biases present significant challenges, meaningful insights into the EOS can be obtained using quantitative corrections and statistics from a large number of simulated BNSs. Addressing these biases will require further efforts in detector technology, waveform modeling, and data analysis methodologies. The ongoing and future observations of BNS mergers will continue to be a critical avenue for probing the fundamental physics of dense matter and understanding neutron stars.

\section{Acknowledgments}
\begin{acknowledgments}
We thank the internal reviewers from the LIGO Scientific Collaboration's Presentation and Publication team and the Virgo and KAGRA collaborations for their feedback on this manuscript. We are grateful for the computational resources provided by LIGO Laboratory and the Leonard E Parker Center for Gravitation, Cosmology and Astrophysics at the University of Wisconsin-Milwaukee and supported by National Science Foundation Grants PHY-0757058, PHY-0823459, PHY-1626190, and PHY-1700765. We especially thank the computing resources provided by Digital Research Alliance of Canada through two consecutive grants, the DRAC RPP $\#$1012 and the Compute Canada Allocation Award $\#$696 to the University of British Columbia. We also thank the Hawk supercomputing system provided by Cardiff University. This material is based upon work supported by NSF's LIGO Laboratory which is a major facility fully funded by the National Science Foundation. AB is founded by the Centre National de la Recherche Scientifique (CNRS). P.L. is supported by the Natural Sciences \& Engineering Research Council of Canada (NSERC). Research at Perimeter Institute is supported in part by the Government of Canada through the Department of Innovation, Science and Economic Development and by the Province of Ontario through the Ministry of Colleges and Universities. HQ was in part supported by the NSF grant PHY-1806656 at Louisiana State University. HQ thanks Jocelyn Read and Vivien Raymond for useful discussions when this collaboration was initially formed.  
\end{acknowledgments}

\bibliography{main.bib}{}

\begin{thebibliography}{}
\expandafter\ifx\csname natexlab\endcsname\relax\def\natexlab#1{#1}\fi
\providecommand{\url}[1]{\href{#1}{#1}}
\providecommand{\dodoi}[1]{doi:~\href{http://doi.org/#1}{\nolinkurl{#1}}}
\providecommand{\doeprint}[1]{\href{http://ascl.net/#1}{\nolinkurl{http://ascl.net/#1}}}
\providecommand{\doarXiv}[1]{\href{https://arxiv.org/abs/#1}{\nolinkurl{https://arxiv.org/abs/#1}}}

\bibitem[{{Aasi} {et~al.}(2015){Aasi}, {Abbott}, {Abbott}, {Abbott},
  {Abernathy}, {Ackley}, {Adams}, {Adams}, {Addesso}, \& et~al.}]{Aasi2015}
{Aasi}, J., {Abbott}, B.~P., {Abbott}, R., {et~al.} 2015, CQGra, 32, 074001,
  \dodoi{10.1088/0264-9381/32/7/074001}

\bibitem[{{Abac} {et~al.}(2024){Abac}, {Abbott}, {Abouelfettouh}, {Acernese},
  {Ackley}, {Adhicary}, {Adhikari}, {Adhikari}, {Adkins}, {Agarwal}, \&
  et~al.}]{GW230529}
{Abac}, A.~G., {Abbott}, R., {Abouelfettouh}, I., {et~al.} 2024, ApJL, 970,
  L34, \dodoi{10.3847/2041-8213/ad5beb}

\bibitem[{Abbott {et~al.}(2016)}]{KAGRA:2013rdx}
Abbott, B.~P., {et~al.} 2016, Living Rev. Rel., 19, 1,
  \dodoi{10.1007/s41114-020-00026-9}

\bibitem[{{Abbott} {et~al.}(2017{\natexlab{a}}){Abbott}, {Abbott}, {Abbott},
  {Acernese}, {Ackley}, {Adams}, {Adams}, {Addesso}, {Adhikari}, {Adya}, \&
  et~al.}]{GW170817}
{Abbott}, B.~P., {Abbott}, R., {Abbott}, T.~D., {et~al.} 2017{\natexlab{a}},
  PhRvL, 119, 161101, \dodoi{10.1103/PhysRevLett.119.161101}

\bibitem[{{Abbott} {et~al.}(2017{\natexlab{b}}){Abbott}, {Abbott}, {Abbott},
  {Acernese}, {Ackley}, {Adams}, {Adams}, {Addesso}, {Adhikari}, {Adya}, \&
  et~al.}]{GW170817_mma}
---. 2017{\natexlab{b}}, ApJL, 848, L12, \dodoi{10.3847/2041-8213/aa91c9}

\bibitem[{{Abbott} {et~al.}(2018{\natexlab{a}}){Abbott}, {Abbott}, {Abbott},
  {Acernese}, {Ackley}, {Adams}, {Adams}, {Addesso}, {Adhikari}, {Adya}, \&
  et~al.}]{GW170817_eos}
---. 2018{\natexlab{a}}, PhRvL, 121, 161101,
  \dodoi{10.1103/PhysRevLett.121.161101}

\bibitem[{{Abbott} {et~al.}(2018{\natexlab{b}}){Abbott}, {Abbott}, {Abbott},
  {Abernathy}, {Acernese}, {Ackley}, {Adams}, {Adams}, {Addesso}, {Adhikari},
  \& et~al.}]{ObservingScenarios}
---. 2018{\natexlab{b}}, LRR, 21, 3, \dodoi{10.1007/s41114-018-0012-9}

\bibitem[{{Abbott} {et~al.}(2019){Abbott}, {Abbott}, {Abbott}, {Acernese},
  {Ackley}, {Adams}, {Adams}, {Addesso}, {Adhikari}, {Adya}, \&
  et~al.}]{GW170817_source}
---. 2019, PhRvX, 9, 011001, \dodoi{10.1103/PhysRevX.9.011001}

\bibitem[{Abbott {et~al.}(2019)}]{LIGOScientific:2018mvr}
Abbott, B.~P., {et~al.} 2019, Phys. Rev. X, 9, 031040,
  \dodoi{10.1103/PhysRevX.9.031040}

\bibitem[{{Abbott} {et~al.}(2020{\natexlab{a}}){Abbott}, {Abbott}, {Abbott},
  {Abraham}, {Acernese}, {Ackley}, {Adams}, {Adya}, {Affeldt}, {Agathos}, \&
  et~al.}]{GW170817_models}
{Abbott}, B.~P., {Abbott}, R., {Abbott}, T.~D., {et~al.} 2020{\natexlab{a}},
  CQGra, 37, 045006, \dodoi{10.1088/1361-6382/ab5f7c}

\bibitem[{{Abbott} {et~al.}(2020{\natexlab{b}}){Abbott}, {Abbott}, {Abbott},
  {Abraham}, {Acernese}, {Ackley}, {Adams}, {Adhikari}, {Adya}, {Affeldt}, \&
  et~al.}]{GW190425}
---. 2020{\natexlab{b}}, ApJL, 892, L3, \dodoi{10.3847/2041-8213/ab75f5}

\bibitem[{{Abbott} {et~al.}(2021){Abbott}, {Abbott}, {Abraham}, {Acernese},
  {Ackley}, {Adams}, {Adams}, {Adhikari}, {Adya}, {Affeldt}, \& et~al.}]{NSBHs}
{Abbott}, R., {Abbott}, T.~D., {Abraham}, S., {et~al.} 2021, ApJL, 915, L5,
  \dodoi{10.3847/2041-8213/ac082e}

\bibitem[{Abbott {et~al.}(2021)}]{LIGOScientific:2020ibl}
Abbott, R., {et~al.} 2021, Phys. Rev. X, 11, 021053,
  \dodoi{10.1103/PhysRevX.11.021053}

\bibitem[{{Abbott} {et~al.}(2023){Abbott}, {Abbott}, {Acernese}, {Ackley},
  {Adams}, {Adhikari}, {Adhikari}, {Adya}, {Affeldt}, {Agarwal}, \&
  et~al.}]{O3bPop}
{Abbott}, R., {Abbott}, T.~D., {Acernese}, F., {et~al.} 2023, PhRvX, 13,
  011048, \dodoi{10.1103/PhysRevX.13.011048}

\bibitem[{Abbott {et~al.}(2023)}]{KAGRA:2021vkt}
Abbott, R., {et~al.} 2023, Phys. Rev. X, 13, 041039,
  \dodoi{10.1103/PhysRevX.13.041039}

\bibitem[{{Acernese} {et~al.}(2015){Acernese}, {Agathos}, {Agatsuma}, {Aisa},
  {Allemandou}, {Allocca}, {Amarni}, {Astone}, {Balestri}, {Ballardin}, \&
  et~al.}]{AcerneseAgathos2015}
{Acernese}, F., {Agathos}, M., {Agatsuma}, K., {et~al.} 2015, CQGra, 32,
  024001, \dodoi{10.1088/0264-9381/32/2/024001}

\bibitem[{{Adhikari} {et~al.}(2021){Adhikari}, {Albataineh}, {Androic},
  {Aniol}, {Armstrong}, {Averett}, {Ayerbe Gayoso}, {Barcus}, {Bellini},
  {Beminiwattha}, \& et~al.}]{AdhikariAlbataineh2021}
{Adhikari}, D., {Albataineh}, H., {Androic}, D., {et~al.} 2021, PhRvL, 126,
  172502, \dodoi{10.1103/PhysRevLett.126.172502}

\bibitem[{{Adhikari} {et~al.}(2020){Adhikari}, {Arai}, {Brooks}, {Wipf},
  {Aguiar}, {Altin}, {Barr}, {Barsotti}, {Bassiri}, {Bell}, \&
  et~al.}]{AdhikariArai2020}
{Adhikari}, R.~X., {Arai}, K., {Brooks}, A.~F., {et~al.} 2020, CQGra, 37,
  165003, \dodoi{10.1088/1361-6382/ab9143}

\bibitem[{Agathos {et~al.}(2015)Agathos, Meidam, Del~Pozzo, Li, Tompitak,
  Veitch, Vitale, \& Van Den~Broeck}]{Agathos:2015uaa}
Agathos, M., Meidam, J., Del~Pozzo, W., {et~al.} 2015, Phys. Rev. D, 92,
  023012, \dodoi{10.1103/PhysRevD.92.023012}

\bibitem[{{Akutsu} {et~al.}(2021){Akutsu}, {Ando}, {Arai}, {Arai}, {Araki},
  {Araya}, {Aritomi}, {Aso}, {Bae}, {Bae}, \& et~al.}]{AkutsuAndo2021}
{Akutsu}, T., {Ando}, M., {Arai}, K., {et~al.} 2021, PTEP, 2021, 05A101,
  \dodoi{10.1093/ptep/ptaa125}

\bibitem[{{Antoniadis} {et~al.}(2013){Antoniadis}, {Freire}, {Wex}, {Tauris},
  {Lynch}, {van Kerkwijk}, {Kramer}, {Bassa}, {Dhillon}, {Driebe}, \&
  et~al.}]{AntoniadisFreire2013}
{Antoniadis}, J., {Freire}, P. C.~C., {Wex}, N., {et~al.} 2013, Sci, 340, 448,
  \dodoi{10.1126/science.1233232}

\bibitem[{{Ashton} {et~al.}(2019){Ashton}, {H{\"u}bner}, {Lasky}, {Talbot},
  {Ackley}, {Biscoveanu}, {Chu}, {Divakarla}, {Easter}, {Goncharov}, \&
  et~al.}]{AshtonHubner2019}
{Ashton}, G., {H{\"u}bner}, M., {Lasky}, P.~D., {et~al.} 2019, ApJS, 241, 27,
  \dodoi{10.3847/1538-4365/ab06fc}

\bibitem[{{Baym} {et~al.}(2019){Baym}, {Furusawa}, {Hatsuda}, {Kojo}, \&
  {Togashi}}]{BaymFurusawa2019}
{Baym}, G., {Furusawa}, S., {Hatsuda}, T., {Kojo}, T., \& {Togashi}, H. 2019,
  ApJ, 885, 42, \dodoi{10.3847/1538-4357/ab441e}

\bibitem[{{Biscoveanu} {et~al.}(2023){Biscoveanu}, {Landry}, \&
  {Vitale}}]{BiscoveanuLandry2023}
{Biscoveanu}, S., {Landry}, P., \& {Vitale}, S. 2023, MNRAS, 518, 5298,
  \dodoi{10.1093/mnras/stac3052}

\bibitem[{{Canizares} {et~al.}(2015){Canizares}, {Field}, {Gair}, {Raymond},
  {Smith}, \& {Tiglio}}]{CanizaresField2015}
{Canizares}, P., {Field}, S.~E., {Gair}, J., {et~al.} 2015, PhRvL, 114, 071104,
  \dodoi{10.1103/PhysRevLett.114.071104}

\bibitem[{{Capano} {et~al.}(2020){Capano}, {Tews}, {Brown}, {Margalit}, {De},
  {Kumar}, {Brown}, {Krishnan}, \& {Reddy}}]{CapanoTews2020}
{Capano}, C.~D., {Tews}, I., {Brown}, S.~M., {et~al.} 2020, NatAs, 4, 625,
  \dodoi{10.1038/s41550-020-1014-6}

\bibitem[{{Chen} {et~al.}(2021){Chen}, {Holz}, {Miller}, {Evans}, {Vitale}, \&
  {Creighton}}]{ChenHolz2021}
{Chen}, H.-Y., {Holz}, D.~E., {Miller}, J., {et~al.} 2021, CQGra, 38, 055010,
  \dodoi{10.1088/1361-6382/abd594}

\bibitem[{{De} {et~al.}(2018){De}, {Finstad}, {Lattimer}, {Brown}, {Berger}, \&
  {Biwer}}]{DeFinstad2018}
{De}, S., {Finstad}, D., {Lattimer}, J.~M., {et~al.} 2018, PhRvL, 121, 091102,
  \dodoi{10.1103/PhysRevLett.121.091102}

\bibitem[{{Dietrich} {et~al.}(2019){Dietrich}, {Samajdar}, {Khan},
  {Johnson-McDaniel}, {Dudi}, \& {Tichy}}]{DietrichSamajdar2019}
{Dietrich}, T., {Samajdar}, A., {Khan}, S., {et~al.} 2019, PhRvD, 100, 044003,
  \dodoi{10.1103/PhysRevD.100.044003}

\bibitem[{{Drischler} {et~al.}(2017){Drischler}, {Hebeler}, \&
  {Schwenk}}]{DrischlerHebeler2017}
{Drischler}, C., {Hebeler}, K., \& {Schwenk}, A. 2017, arXiv, arXiv:1710.08220,
  \dodoi{10.48550/arXiv.1710.08220}

\bibitem[{{Essick} {et~al.}(2024){Essick}, {Landry}, {Chatziioannou}, {Legred},
  \& {Ng}}]{lwp2024}
{Essick}, R., {Landry}, P., {Chatziioannou}, K., {Legred}, I., \& {Ng}, S.
  2024, lwp

\bibitem[{{Essick} {et~al.}(2020){Essick}, {Landry}, \&
  {Holz}}]{EssickLandry2020}
{Essick}, R., {Landry}, P., \& {Holz}, D.~E. 2020, PhRvD, 101, 063007,
  \dodoi{10.1103/PhysRevD.101.063007}

\bibitem[{{Evans} {et~al.}(2021){Evans}, {Adhikari}, {Afle}, {Ballmer},
  {Biscoveanu}, {Borhanian}, {Brown}, {Chen}, {Eisenstein}, {Gruson}, \&
  et~al.}]{EvansAdhikari2021}
{Evans}, M., {Adhikari}, R.~X., {Afle}, C., {et~al.} 2021, arXiv,
  arXiv:2109.09882, \dodoi{10.48550/arXiv.2109.09882}

\bibitem[{{Flanagan} \& {Hinderer}(2008)}]{FlanaganHinderer2008}
{Flanagan}, {\'E}.~{\'E}., \& {Hinderer}, T. 2008, PhRvD, 77, 021502,
  \dodoi{10.1103/PhysRevD.77.021502}

\bibitem[{{Fonseca} {et~al.}(2021){Fonseca}, {Cromartie}, {Pennucci}, {Ray},
  {Kirichenko}, {Ransom}, {Demorest}, {Stairs}, {Arzoumanian}, {Guillemot}, \&
  et~al.}]{FonsecaCromartie2021}
{Fonseca}, E., {Cromartie}, H.~T., {Pennucci}, T.~T., {et~al.} 2021, ApJL, 915,
  L12, \dodoi{10.3847/2041-8213/ac03b8}

\bibitem[{Ghosh {et~al.}(2024)Ghosh, Biswas, Bose, \& Kapadia}]{Ghosh:2024cwc}
Ghosh, T., Biswas, B., Bose, S., \& Kapadia, S.~J. 2024.
\newblock \doarXiv{2407.16669}

\bibitem[{Hernandez~Vivanco {et~al.}(2019)Hernandez~Vivanco, Smith, Thrane,
  Lasky, Talbot, \& Raymond}]{HernandezVivanco:2019vvk}
Hernandez~Vivanco, F., Smith, R., Thrane, E., {et~al.} 2019, Phys. Rev. D, 100,
  103009, \dodoi{10.1103/PhysRevD.100.103009}

\bibitem[{{Huth} {et~al.}(2022){Huth}, {Pang}, {Tews}, {Dietrich}, {Le
  F{\`e}vre}, {Schwenk}, {Trautmann}, {Agarwal}, {Bulla}, {Coughlin}, \&
  et~al.}]{HuthPang2022}
{Huth}, S., {Pang}, P. T.~H., {Tews}, I., {et~al.} 2022, Natur, 606, 276,
  \dodoi{10.1038/s41586-022-04750-w}

\bibitem[{Kiziltan {et~al.}(2013)Kiziltan, Kottas, De~Yoreo, \&
  Thorsett}]{Kiziltan:2013oja}
Kiziltan, B., Kottas, A., De~Yoreo, M., \& Thorsett, S.~E. 2013, Astrophys. J.,
  778, 66, \dodoi{10.1088/0004-637X/778/1/66}

\bibitem[{{Koehn} {et~al.}(2024){Koehn}, {Rose}, {Pang}, {Somasundaram},
  {Reed}, {Tews}, {Abac}, {Komoltsev}, {Kunert}, {Kurkela}, \&
  et~al.}]{KoehnRose2024}
{Koehn}, H., {Rose}, H., {Pang}, P. T.~H., {et~al.} 2024, arXiv,
  arXiv:2402.04172, \dodoi{10.48550/arXiv.2402.04172}

\bibitem[{{Landry} \& {Essick}(2019)}]{LandryEssick2019}
{Landry}, P., \& {Essick}, R. 2019, PhRvD, 99, 084049,
  \dodoi{10.1103/PhysRevD.99.084049}

\bibitem[{Landry {et~al.}(2020)Landry, Essick, \&
  Chatziioannou}]{Landry:2020vaw}
Landry, P., Essick, R., \& Chatziioannou, K. 2020, Phys. Rev. D, 101, 123007,
  \dodoi{10.1103/PhysRevD.101.123007}

\bibitem[{{Landry} {et~al.}(2020){Landry}, {Essick}, \&
  {Chatziioannou}}]{LandryEssick2020}
{Landry}, P., {Essick}, R., \& {Chatziioannou}, K. 2020, PhRvD, 101, 123007,
  \dodoi{10.1103/PhysRevD.101.123007}

\bibitem[{{Landry} \& {Poisson}(2014)}]{LandryPoisson2014}
{Landry}, P., \& {Poisson}, E. 2014, PhRvD, 89, 124011,
  \dodoi{10.1103/PhysRevD.89.124011}

\bibitem[{{Landry} \& {Read}(2021)}]{LandryRead2021}
{Landry}, P., \& {Read}, J.~S. 2021, ApJL, 921, L25,
  \dodoi{10.3847/2041-8213/ac2f3e}

\bibitem[{{Le F{\`e}vre} {et~al.}(2016){Le F{\`e}vre}, {Leifels}, {Reisdorf},
  {Aichelin}, \& {Hartnack}}]{LeFevreLeifels2016}
{Le F{\`e}vre}, A., {Leifels}, Y., {Reisdorf}, W., {Aichelin}, J., \&
  {Hartnack}, C. 2016, NuPhA, 945, 112, \dodoi{10.1016/j.nuclphysa.2015.09.015}

\bibitem[{{Legred} {et~al.}(2021){Legred}, {Chatziioannou}, {Essick}, {Han}, \&
  {Landry}}]{LegredChatziioannou2021}
{Legred}, I., {Chatziioannou}, K., {Essick}, R., {Han}, S., \& {Landry}, P.
  2021, PhRvD, 104, 063003, \dodoi{10.1103/PhysRevD.104.063003}

\bibitem[{{Lynn} {et~al.}(2016){Lynn}, {Tews}, {Carlson}, {Gandolfi},
  {Gezerlis}, {Schmidt}, \& {Schwenk}}]{LynnTews2016}
{Lynn}, J.~E., {Tews}, I., {Carlson}, J., {et~al.} 2016, PhRvL, 116, 062501,
  \dodoi{10.1103/PhysRevLett.116.062501}

\bibitem[{{Maggiore} {et~al.}(2020){Maggiore}, {Van Den Broeck}, {Bartolo},
  {Belgacem}, {Bertacca}, {Bizouard}, {Branchesi}, {Clesse}, {Foffa},
  {Garc{\'\i}a-Bellido}, \& et~al.}]{MaggioreVanDenBroeck2020}
{Maggiore}, M., {Van Den Broeck}, C., {Bartolo}, N., {et~al.} 2020, JCAP, 2020,
  050, \dodoi{10.1088/1475-7516/2020/03/050}

\bibitem[{{Miller} {et~al.}(2019){Miller}, {Lamb}, {Dittmann}, {Bogdanov},
  {Arzoumanian}, {Gendreau}, {Guillot}, {Harding}, {Ho}, {Lattimer}, \&
  et~al.}]{MillerLamb2019}
{Miller}, M.~C., {Lamb}, F.~K., {Dittmann}, A.~J., {et~al.} 2019, ApJL, 887,
  L24, \dodoi{10.3847/2041-8213/ab50c5}

\bibitem[{{Miller} {et~al.}(2021){Miller}, {Lamb}, {Dittmann}, {Bogdanov},
  {Arzoumanian}, {Gendreau}, {Guillot}, {Ho}, {Lattimer}, {Loewenstein}, \&
  et~al.}]{MillerLamb2021}
---. 2021, ApJL, 918, L28, \dodoi{10.3847/2041-8213/ac089b}

\bibitem[{Morisaki {et~al.}(2023)Morisaki, Smith, Tsukada, Sachdev, Stevenson,
  Talbot, \& Zimmerman}]{PhysRevD.108.123040}
Morisaki, S., Smith, R., Tsukada, L., {et~al.} 2023, Phys. Rev. D, 108, 123040,
  \dodoi{10.1103/PhysRevD.108.123040}

\bibitem[{Mould {et~al.}(2024)Mould, Moore, \& Gerosa}]{Mould:2023eca}
Mould, M., Moore, C.~J., \& Gerosa, D. 2024, Phys. Rev. D, 109, 063013,
  \dodoi{10.1103/PhysRevD.109.063013}

\bibitem[{{M{\"u}ther} {et~al.}(1987){M{\"u}ther}, {Prakash}, \&
  {Ainsworth}}]{MutherPrakash1987}
{M{\"u}ther}, H., {Prakash}, M., \& {Ainsworth}, T.~L. 1987, PhLB, 199, 469,
  \dodoi{10.1016/0370-2693(87)91611-X}

\bibitem[{{Oppenheimer} \& {Volkoff}(1939)}]{OppenheimerVolkoff1939}
{Oppenheimer}, J.~R., \& {Volkoff}, G.~M. 1939, PhRv, 55, 374,
  \dodoi{10.1103/PhysRev.55.374}

\bibitem[{Ozel {et~al.}(2012)Ozel, Psaltis, Narayan, \&
  Villarreal}]{Ozel:2012ax}
Ozel, F., Psaltis, D., Narayan, R., \& Villarreal, A.~S. 2012, Astrophys. J.,
  757, 55, \dodoi{10.1088/0004-637X/757/1/55}

\bibitem[{Pradhan {et~al.}(2024)Pradhan, Ghosh, Pathak, \&
  Chatterjee}]{Pradhan:2023xtq}
Pradhan, B.~K., Ghosh, T., Pathak, D., \& Chatterjee, D. 2024, Astrophys. J.,
  966, 79, \dodoi{10.3847/1538-4357/ad31a8}

\bibitem[{Qi \& Raymond(2021)}]{PhysRevD.104.063031}
Qi, H., \& Raymond, V. 2021, Phys. Rev. D, 104, 063031,
  \dodoi{10.1103/PhysRevD.104.063031}

\bibitem[{{Raaijmakers} {et~al.}(2019){Raaijmakers}, {Riley}, {Watts}, {Greif},
  {Morsink}, {Hebeler}, {Schwenk}, {Hinderer}, {Nissanke}, {Guillot}, \&
  et~al.}]{RaaijmakersRiley2019}
{Raaijmakers}, G., {Riley}, T.~E., {Watts}, A.~L., {et~al.} 2019, ApJL, 887,
  L22, \dodoi{10.3847/2041-8213/ab451a}

\bibitem[{{Reinhard} \& {Flocard}(1995)}]{ReinhardFlocard1995}
{Reinhard}, P.~G., \& {Flocard}, H. 1995, NuPhA, 584, 467,
  \dodoi{10.1016/0375-9474(94)00770-N}

\bibitem[{{Russotto} {et~al.}(2016){Russotto}, {Gannon}, {Kupny}, {Lasko},
  {Acosta}, {Adamczyk}, {Al-Ajlan}, {Al-Garawi}, {Al-Homaidhi}, {Amorini}, \&
  et~al.}]{RussottoGannon2016}
{Russotto}, P., {Gannon}, S., {Kupny}, S., {et~al.} 2016, PhRvC, 94, 034608,
  \dodoi{10.1103/PhysRevC.94.034608}

\bibitem[{{Salmi} {et~al.}(2024){Salmi}, {Choudhury}, {Kini}, {Riley},
  {Vinciguerra}, {Watts}, {Wolff}, {Arzoumanian}, {Bogdanov}, {Chakrabarty}, \&
  et~al.}]{SalmiChoudhury2024}
{Salmi}, T., {Choudhury}, D., {Kini}, Y., {et~al.} 2024, arXiv,
  arXiv:2406.14466, \dodoi{10.48550/arXiv.2406.14466}

\bibitem[{Sarin {et~al.}(2024)Sarin, Peiris, Mortlock, Alsing, Nissanke, \&
  Feeney}]{Sarin:2023tgv}
Sarin, N., Peiris, H.~V., Mortlock, D.~J., {et~al.} 2024, Phys. Rev. D, 110,
  024076, \dodoi{10.1103/PhysRevD.110.024076}

\bibitem[{{Smith} {et~al.}(2016){Smith}, {Field}, {Blackburn}, {Haster},
  {P{\"u}rrer}, {Raymond}, \& {Schmidt}}]{SmithField2016}
{Smith}, R., {Field}, S.~E., {Blackburn}, K., {et~al.} 2016, PhRvD, 94, 044031,
  \dodoi{10.1103/PhysRevD.94.044031}

\bibitem[{T2000012-v2(2022)}]{LVKcurvesAdv}
T2000012-v2. 2022, Noise curves used for Simulations in the update of the
  Observing Scenarios Paper

\bibitem[{\text{Shoemaker et al.}(2024)}]{LVKobsplan}
\text{Shoemaker et al.} 2024, {LIGO, Virgo, and KAGRA Observing Run Plans},
  \url{https://observing.docs.ligo.org/plan}

\bibitem[{{Tolman}(1939)}]{Tolman1939}
{Tolman}, R.~C. 1939, PhRv, 55, 364, \dodoi{10.1103/PhysRev.55.364}

\bibitem[{{Wade} {et~al.}(2014){Wade}, {Creighton}, {Ochsner}, {Lackey},
  {Farr}, {Littenberg}, \& {Raymond}}]{WadeCreighton2014}
{Wade}, L., {Creighton}, J. D.~E., {Ochsner}, E., {et~al.} 2014, PhRvD, 89,
  103012, \dodoi{10.1103/PhysRevD.89.103012}

\bibitem[{{Wysocki} {et~al.}(2020){Wysocki}, {O'Shaughnessy}, {Wade}, \&
  {Lange}}]{WysockiO'Shaughnessy2020}
{Wysocki}, D., {O'Shaughnessy}, R., {Wade}, L., \& {Lange}, J. 2020, arXiv,
  arXiv:2001.01747, \dodoi{10.48550/arXiv.2001.01747}

\end{thebibliography}
\bibliographystyle{aasjournal}



\end{document}